# On the Reliability of Sampling Strategies in Offline Recommender Evaluation

Bruno L. Pereira
Universidade Federal de Minas Gerais
Belo Horizonte, MG, Brazil
brunolaporais@dcc.ufmg.br

Alan Said
University of Gothenburg
Gothenburg, Sweden
alansaid@acm.org

Rodrygo L. T. Santos
Universidade Federal de Minas Gerais
Belo Horizonte, MG, Brazil
rodrygo@dcc.ufmg.br

## Abstract

Offline evaluation plays a central role in benchmarking recommender systems when online testing is impractical or risky. However, it is susceptible to two key sources of bias: exposure bias, where users only interact with items they are shown, and sampling bias, introduced when evaluation is performed on a subset of logged items rather than the full catalog. While prior work has proposed methods to mitigate sampling bias, these are typically assessed on fixed logged datasets rather than for their ability to support reliable model comparisons under varying exposure conditions or relative to true user preferences. In this paper, we investigate how different combinations of logging and sampling choices affect the reliability of offline evaluation. Using a fully observed dataset as ground truth, we systematically simulate diverse exposure biases and assess the reliability of common sampling strategies along four dimensions: sampling resolution (recommender model separability), fidelity (agreement with full evaluation), robustness (stability under exposure bias), and predictive power (alignment with ground truth). Our findings highlight when and how sampling distorts evaluation outcomes and offer practical guidance for selecting strategies that yield faithful and robust offline comparisons.

## 1 Introduction

In an ideal world, recommender systems would be evaluated by observing how users respond to fully ranked lists of items, ensuring that every item has a fair chance of being chosen [15, 23]. While this is rarely achievable in practice, online A/B testing allows for more controlled exposure, where interaction data more reliably reflects user preferences over the presented items, and exposure bias can be actively managed by the experimenter. However, such tests are costly and risky: researchers may lack access to live traffic, and practitioners may hesitate to expose users to untested models. As a result, offline evaluation remains a prevalent practice, despite its well-documented limitations [7, 22].

A well-known issue in offline evaluation is selection bias, also referred to as logging bias [48]. Since users interact only with items they were shown, the resulting data is often extremely sparse [46] and *missing-not-at-random* (MNAR) [35]. This skews evaluation in favor of models that replicate past exposure patterns rather than uncovering genuine preferences [39, 42]. While various techniques have been proposed to mitigate this—such as inverse propensity scoring [45] and counterfactual estimators [17]—this paper focuses on a second, often overlooked source of distortion: *sampling bias*.

Due to the computational cost incurred by ranking over large item catalogs, most offline evaluation pipelines sample a small set of negatives for each positive instance when computing top-$k$ metrics. The way these negatives are sampled—whether uniformly [10, 24], by popularity [6, 11], by positivity [39], or using debiasing strategies like weighted sampling (WTD) [5] or Skew [21, 30]—can substantially impact evaluation outcomes [8, 25, 27, 33]. However, prior work typically assesses sampling strategies using fixed logged datasets, without accounting for different sources or intensities of exposure bias [15, 21, 51]. Moreover, most studies take the logged data itself as ground truth, leaving open the question of how well sampled evaluation outcomes reflect actual user preferences.

To systematically examine these effects, we distinguish between the data we wish we had—fully observed user preferences—and the data actually available for evaluation, which is shaped by logging and sampling decisions. Inspired by past research, we investigate the reliability of existing sampling strategies across a range of sample sizes [15, 21, 51]. Unlike prior studies on sampling strategies, we conduct experiments under multiple simulated exposure bias conditions and provide the first systematic analysis of how these sampling strategies perform—not just against biased logged data, but also in comparison to ground-truth user preferences. By disentangling the effects of exposure and sampling bias, this work establishes a more principled foundation for offline evaluation and offers practical guidance on selecting sampling strategies in the presence of biased data. Our contributions are as follows:

- We formalize four dimensions of sampling reliability: resolution (model separability), fidelity (agreement with full evaluation), robustness (stability under bias), and predictive power (alignment with ground truth).
- We propose an evaluation framework based on a fully observed dataset, enabling a controlled assessment of sampled evaluation under both unbiased and biased exposure.
- We conduct a comprehensive empirical analysis of widely used sampling strategies across multiple logging configurations, assessing their reliability in ranking-based evaluation.

In the remainder of this paper, Section 2 reviews related work on exposure bias, sampling strategies, and the reliability of evaluation protocols. Section 3 introduces our conceptual framework for sampling reliability assessment. Section 4 presents our experimental setup, including how we simulate exposure bias ahead of sampling. Section 5 reports our results, structured around four dimensions of evaluation reliability. Finally, Section 6 concludes with practical implications and directions for future work.

## 2 Related Work

A rich body of research has examined the challenges of offline recommender evaluation, particularly in the presence of biases introduced by system-driven exposure and incomplete observations. In this section, we review related work in three areas: methods for mitigating exposure bias in logged data, approaches to addressing sampling bias in evaluation protocols, and recent efforts focused on assessing the reliability of evaluation procedures.

### 2.1 Exposure Bias in Logged Data

User interaction data in recommender systems is inherently biased by the system's exposure policies. Users can only interact with items they are shown, which means that logged interactions are confounded by both user preferences and the system's exposure decisions [48]. This phenomenon—known as *exposure bias*—renders the logged data MNAR and undermines the validity of both learning and evaluation [9]. Prior work has characterized the nature of exposure bias [4] and proposed several mitigation strategies. These include inverse propensity weighting and its variants [41, 45, 47], counterfactual learning approaches [12, 18, 20, 29, 31], and self-supervised methods for estimating propensities [32]. Some studies adopt doubly robust estimation to combine reward modeling with exposure modeling for improved accuracy [13].

While much of this literature focuses on learning unbiased recommenders, a growing number of works examine how exposure bias affects evaluation. Wang et al. [51] and Liu et al. [34] show that biased logging policies can significantly distort offline evaluation metrics and relative model rankings, leading to potentially misleading conclusions. Other studies have examined the implications for fairness [1, 19], highlighting how skewed exposure can amplify disparities across users or item groups. Several works propose adjusted evaluation procedures that assume missing-at-random (MAR) exposure and seek to correct standard metrics accordingly [35, 51]. Finally, randomized exposure datasets such as Yahoo R3 [35] and KuaiRand [16] provide empirical testbeds for enabling more principled evaluation, helping to isolate the effects of specific exposure policies from underlying user preferences.

### 2.2 Sampling Bias in Offline Evaluation

Offline evaluation is also shaped by sampling candidate items used to approximate top-$k$ metrics. Because scoring all items is computationally expensive, evaluations typically sample a subset of non-positive items for each user. This introduces *sampling bias*, which can significantly affect evaluation metric values and, in turn, alter the relative ranking of models by over- or underestimating their true recommendation performance [8, 21, 25, 27].

The choice of sampling strategy matters. Sampling uniformly or by popularity [27], positivity [33], or exposure [6] can each lead to different evaluation outcomes. Some methods aim to approximate the full ranking using weighted correction techniques [6, 30], or adaptively adjust the number of items sampled per user [28]. An alternative to sampling unexposed items as synthetic negatives is to evaluate over impressions—items that were actually exposed to the user, regardless of interaction. This naturally avoids unexposed false negatives and grounds evaluation in observed choice sets [37]. Impression-based evaluation has been explored in datasets like ContentWise [38], MIND [52], and FINN.no [14]—but such datasets are rare, and the reliability of impression-based metrics under different exposure biases remains underexplored. In our study, we extend this line of work by explicitly assessing how different strategies for sampling exposure data affect evaluation outcomes.

### 2.3 Reliability of Evaluation Protocols

A third line of work examines the broader question of whether offline evaluation protocols reliably reflect true model performance—a concern known as *meta-evaluation*. This perspective emphasizes that even well-intentioned evaluation strategies may yield misleading conclusions if confounding factors are unaccounted for.

In information retrieval, Sanderson et al. [43] introduced the concept of *predictive power*, defined as the degree to which partial judgments (e.g. sampled relevance) recover system rankings derived from complete judgments. In recommendation, Krichene and Rendle [25] showed that commonly used sampled metrics like recall@$k$ can be unfaithful estimators of ranking performance when evaluation is conducted over sampled candidates. Liu et al. [34] applied item response theory to model the latent abilities of recommender algorithms and the difficulty of recommendation tasks, revealing that high scores on rank-based metrics often fail to reflect genuine improvements in modeling user preferences.

Other studies have focused on the evaluation sensitivity to design choices. Carraro and Bridge [6] and Cañamares and Castells [8] conducted empirical evaluations of how sampling strategies affect metric consistency. Ihemelandu and Ekstrand [21] provided a detailed framework for analyzing candidate set sampling under known exposure. Surveys such as Castells and Moffat [7] and Zangerle and Bauer [53] have emphasized the fragility of current evaluation practices, highlighting issues such as test set leakage, missing negative feedback, and the lack of robustness across tasks.

Most prior studies on sampling reliability use fixed exposure logs and compare sampled evaluation against full evaluation over those logs [25, 34]. Our work expands on this by systematically varying both the exposure policy and the sampling strategy, and evaluating reliability across four dimensions: *resolution*, the capacity to distinguish between competing models; *fidelity*, the agreement between sampled and full evaluations; *robustness*, the stability of results under varying bias conditions; and *predictive power*, the ability to recover ground-truth model rankings.

## 3 Conceptual Framework

To facilitate systematic analysis, we develop a conceptual framework that separates offline evaluation into three sequential stages: user preferences, item exposure, and evaluation sampling. This stratified approach allows us to formally characterize the discrepancy between ideal evaluation based on complete preference information and practical evaluation constrained by exposure bias and sampling limitations. Let $U$ be the set of users and $I$ the set of items. We define three artifacts central to our evaluation design:

*Ground-Truth Preferences (G).* We assume a fully observed user-item preference matrix $G \in \mathbb{R}^{|U| \times |I|}$, in which each entry indicates the extent to which an item is relevant to a user. This matrix represents an idealized view of user preferences—every user has an opinion about every item. While such complete data is seldom available in practice, $G$ serves as a notional ground truth against which the reliability of evaluation pipelines can be measured.

*Logged Interactions (L).* In real-world systems, users are exposed to only a subset of items. The logged matrix $L$ is a partially observed version of $G$, derived by applying an exposure policy. It contains interactions only for items shown to each user, while unexposed entries remain missing. Thus, $L$ reflects both user preferences and exposure bias introduced by the system's logging mechanism. Without loss of generality, we assume a typical implicit feedback scenario, where observed interactions are interpreted as either positive (indicating meaningful engagement) or not.

*Sampled Interactions ($G_S$ and $L_S$).* To operationalize top-$k$ evaluation, it is common to construct evaluation sets by pairing each user's known positives with a sampled subset of non-positives—that is, items that were either exposed but not interacted with, or not exposed at all. This yields a typical sampled offline evaluation scenario, $L_S$, where sampling is applied to the partially observed logged matrix $L$. To isolate the impact of sampling from exposure effects, we also consider a sibling scenario, $G_S$, where sampling is applied to the fully observed ground-truth matrix $G$.

This conceptual framework centers $L_S$ as the primary evaluation artifact and supports a systematic analysis of multiple sources of distortion. We begin by assessing how well $L_S$ distinguishes among competing recommender models, reflecting evaluation resolution as an intrinsic property of the sampling protocol itself. Next, comparing $L_S$ to $L$ (or $G_S$ to $G$) isolates the effect of sampling under fixed exposure. In turn, comparing $L_S$ to $G_S$ captures the impact of exposure bias under fixed sampling. Finally, comparing $L_S$ to $G$ reveals the combined distortion introduced by both logging and sampling, and ultimately, the extent to which sampled evaluation over biased data reflects true user preferences.

## 4 Experimental Setup

Our experiments are designed to evaluate the reliability of sampling-based offline evaluation across four complementary dimensions. Taking $L_S$ as our main object of study—the artifact produced by applying a sampling strategy to logged data—we assess the quality of the resulting evaluation along the following axes:

*Q1. Resolution: Can the sampler distinguish between models?* We ask whether sampling provides enough resolution to differentiate between recommender systems, measured by the number of ties in model rankings produced under $L_S$.

*Q2. Fidelity: Does the sampler preserve full evaluation outcomes?* We compare sampled evaluations ($L_S$ and $G_S$) to their full counterparts ($L$ and $G$) to assess whether sampling introduces distortions, even when the data source is fixed.

*Q3. Robustness: Is the sampler stable under bias?* We apply the same sampling strategy to both unbiased ground-truth data ($G$) and biased logged data ($L$), and assess whether the evaluation outcomes remain consistent.

*Q4. Predictive Power: Does the sampler help us recover the truth?* We examine the extent to which evaluations based on a biased sample ($L_S$) yield the same model rankings we would obtain from full evaluation on ground-truth preferences ($G$).

The remainder of this section describes how we instantiate the key elements of our framework introduced in Section 3—ground-truth preferences ($G$), logged observations ($L$), and sampled evaluations ($G_S, L_S$)—and how we vary logger policies, sampling strategies, and recommender models to answer these questions. All experiments were conducted by extending the Microsoft Recommenders framework (v1.2.0).[1] Experiments were run on a Linux workstation with Ubuntu 22.04.4 LTS and Python 3.10.16, equipped with an Intel® Xeon® Silver 4314 CPU (16 cores, 32 threads, base clock 2.4 GHz), 512 GB DDR4 RAM, and a single NVIDIA® A100 GPU. To ensure reproducibility, all code is publicly available.[2]

### 4.1 Dataset and Ground Truth Construction

Our experiments use the KuaiRec dataset [15],[3] a large-scale collection of user interactions with short-form videos. As illustrated in Fig. 1, interactions are treated as implicit feedback, with binary relevance labels derived from user engagement behavior. Specifically, following Gao et al. [15], we define positive feedback as cases where the cumulative watch time of a video exceeds twice its duration. The dataset is partitioned into training and test splits following the original authors' methodology. Summary statistics for both splits are provided in Table 1. Notably, the test split contains complete interaction histories (i.e. nearly 100% density), allowing us to construct a fully-observed binary relevance matrix that serves as our ground-truth preference matrix $G$. To the best of our knowledge, KuaiRec is the only publicly available dataset with full exposure logs, enabling the controlled simulations central to our study.

Table 1: KuaiRec dataset statistics.

|  | #Users | #Items | #Interact$^\oplus$ | #Interact$^\ominus$ | Density |
|---|---|---|---|---|---|
| *Train* | 7,176 | 10,728 | 936,390 | 11,594,416 | 16.3% |
| *Test* | 1,411 | 3,327 | 217,175 | 4,459,395 | 99.6% |

---

[1] https://github.com/recommenders-team/recommenders
[2] https://github.com/LatinUFMG/recommenders-sampling
[3] https://kuairec.com/

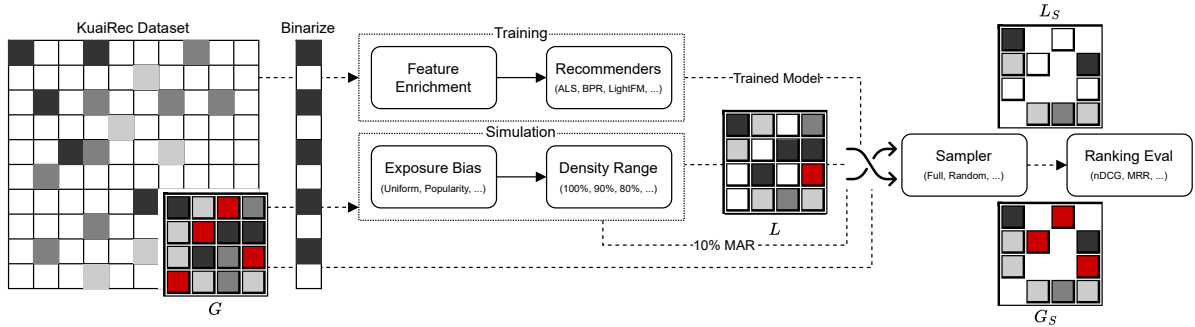

Figure 1: The experimental setup using the KuaiRec dataset.

## 4.2 Logger Simulation

Following Gao et al. [15], to obtain partially observed data $L$, we simulate exposure using three logging strategies applied to $G$:

- **Uniform:** items are exposed to users uniformly at random, reproducing a baseline, unbiased exposure scenario.
- **Popularity-biased:** items are exposed with probabilities proportional to their global popularity in $G$.
- **Positivity-biased:** items are exposed with probabilities proportional to their positive feedback count in $G$.

These logging strategies are inspired by prior work on modeling and analyzing exposure bias in recommendation [21, 30]. For each logger, we retain a subset of each user's interactions from $G$ according to target sparsity levels[4]—0%, 10%, 30%, 50%, 70%, 85%, 90%, 95%—while ensuring at least one positive item per test user. The resulting matrix $L$ captures logged interactions shaped by exposure bias.

## 4.3 Sampling Strategies

From each simulated logged interactions matrix $L$, we construct multiple sampled evaluation matrices $L_S$ by retaining all known positives and appending a set of non-positive items per test user, according to one of the following sampling strategies:

(1) **Full:** uses all available non-positives (no sampling).
(2) **Exposed:** uses all *exposed* non-positives for each user.
(3) **Random@$e$:** samples $e$ non-positives uniformly at random, where $e$ is the number of exposed non-positives per user.
(4) **Random@$n$:** samples $n$ non-positives uniformly at random.
(5) **Popularity@$n$:** samples $n$ non-positive items, where selection probabilities are determined by applying Zipf's law [36] based on each item's popularity rank [15].
(6) **Positivity@$n$:** samples $n$ non-positive items, with selection probabilities weighted by their estimated relevance scores, also following Zipf's law [15, 36].
(7) **WTD@$n$:** samples $n$ non-positives weighted based on empirical exposure distributions [5].

[4] A random 10% of $G$ is held out prior to logging and consistently excluded from all logger simulations. This partition is used exclusively to support samplers that require unbiased data for propensity estimation [5], as discussed in Section 4.3. Sparsity levels here refer to the remainder of $G$ after this holdout.

(8) **WTDH@$n$:** heuristic version of WTD with uniform exposure assumptions.
(9) **Skew@$n$:** samples $n$ non-positives, with selection probabilities defined by empirical popularity distributions [21, 30].

For samplers (4)–(9), we vary the number of non-positives sampled, $n \in \{1, 2, 5, 10, 20, 50, 100, 200, 500, 1000\}$, to assess its impact on the outcome of the evaluation process. Combined with the 24 loggers introduced in Section 4.2 (3 logging strategies × 8 sparsity levels), these 63 samplers—comprising 3 fixed samplers and 6 parametric ones, each evaluated at 10 sample sizes—yield a total of 1,512 evaluation scenarios used in our analysis.

## 4.4 Recommender Models

To assess sampling reliability across diverse recommendation techniques, we implement and evaluate the following models, covering distinct algorithmic families:

- **ALS** [49]: Alternating Least Squares is a matrix factorization technique that models user-item interactions through latent factors. It alternately optimizes user and item embeddings to minimize squared error, and is well-suited for implicit feedback.
- **BPR** [40]: Bayesian Personalized Ranking is a pairwise ranking method that optimizes for item ranking by learning to prefer observed interactions over unobserved ones, using a Bayesian framework.
- **LightFM** [26]: A hybrid model combining collaborative and content-based signals. We train two variants: one using collaborative signals only, and another leveraging metadata features. Its flexibility makes it effective in both sparse and cold-start settings.
- **SAR-Cosine** [44]: A lightweight item-based collaborative filtering approach that computes item-item similarities using cosine similarity.
- **SAR-Jaccard** [44]: A variant of SAR-Cosine that uses the Jaccard index for similarity computation, making it particularly suitable for binary interaction data.
- **Popularity**: A non-personalized baseline that ranks items by global popularity, independent of the target user.
- **Random**: A naive baseline that ranks items randomly.

We optimized hyperparameters for each model using random search [2] with the Hyperopt library[5] and parallelized evaluations via Ray Tune.[6] Each search ran for 128 iterations, with candidate configurations evaluated through 5-fold cross-validation on random splits of the training set. Full details on search ranges and selected values are available in the public code repository.

### 4.5 (Meta-)Evaluation Metrics

Each of the seven models described in Section 4.4 is evaluated on the same sampled matrix $L_S$, for a total of 1,512 such matrices, as detailed in Section 4.3. For each case, we compute precision, recall, and nDCG at various ranking cutoffs (@5, @10, @50, @100). While we report results primarily for nDCG@100 due to space constraints, this deeper cutoff was deliberately chosen to better assess the complete ranking behavior of sampling strategies. Although users typically interact with only the top few items, evaluation sampling affects the entire ranked list. Corroborating the findings of Valcarce et al. [50], our analyses showed that nDCG@100 provides greater resolution between sampling strategies than conventional shorter cutoffs, while still reflecting the same directional patterns observed across all metrics and cutoffs. Based on these evaluation results, we assess the reliability of each $L_S$ construction protocol using two meta-evaluation metrics:

- **Tie rate:** the fraction of tied recommender model pairs in each individual evaluation scenario [8]; used to address *Q1*.
- **Kendall's $\tau$:** the degree of ranking agreement across pairs of evaluation scenarios; used to address *Q2–Q4*.

## 5 Experimental Results

We now present findings from our experimental study, organized around four key dimensions of evaluation reliability. We begin with resolution (*Q1*)—whether different sampling strategies can effectively separate recommender models based on their performance. We then analyze fidelity (*Q2*)—how closely each strategy reproduces model rankings obtained under full evaluation—and robustness (*Q3*)—the stability of these rankings under varying exposure biases. Finally, we address predictive power (*Q4*)—the extent to which evaluation over biased and sampled data recovers the ground-truth ranking defined by fully observed preferences. For all experiments, confidence intervals (95%, from 1,000 bootstrap resamples) were computed but most are too small to be visible.

### 5.1 Q1: Resolution

In this section, we address *Q1* by assessing the extent to which different sampling strategies provide sufficient resolution to distinguish between recommender models. We quantify resolution using the tie rate [3, 8], defined as the fraction of recommender model pairs that achieve identical nDCG@100 scores for a given user. While the tie rate serves as a proxy for discriminative power, it fundamentally differs from interpretations rooted in statistical significance testing. This measure is particularly valuable in scenarios where traditional statistical tests may be underpowered—such as in cases involving very small or very large sample sizes [8]. Lower tie rates indicate greater ability to resolve differences in model quality—hence greater resolution. For each configuration, we compute tie rates at the user level and average them across users.

[5]https://github.com/hyperopt/hyperopt
[6]https://docs.ray.io/en/latest/tune/index.html

Fig. 2 presents the average tie rate (log scale on the $y$-axis) as a function of sample size $n$ (on the $x$-axis), for various sampling strategies (shown as plot lines). Horizontal lines represent fixed-size samplers (Full, Exposed, and Random@$e$), which do not vary with $n$. Each subplot corresponds to a unique combination of logging strategy (rows: Uniform, Popularity-biased, Positivity-biased) and sparsity level (columns, from 0% to 95%).

From Fig. 2, we first note that, in line with previous research [8], tie rates are particularly high in too small or too large samples, when recommenders are more likely to tie at either very high or very low nDCG values, respectively. However, logger sparsity also plays a critical role: under low sparsity levels (e.g., 0–50%), most samplers achieve low tie rates with small to moderate sample sizes, indicating good resolution. As sparsity increases—particularly at 90% and 95%—tie rates remain substantially higher, even for large $n$, due to the loss of signal in the logged data. The choice of sampler becomes especially important under these challenging conditions. Samplers like Skew, Popularity, and Positivity consistently yield lower tie rates, effectively preserving meaningful distinctions between models even under high sparsity and biased exposure. On the other hand, the Full sampler, which includes all items, often yields lower resolution in distinguishing among models across most settings—particularly as sparsity increases. Exposed and Random@$e$ serve as useful baselines, assuming access to full exposure or a fixed sample size based on exposed items, and show improved performance in high-sparsity scenarios. Overall, samplers with small to moderate size $n$ continue to deliver robust results.

Recalling *Q1*, not all samplers are equally capable of distinguishing models. The ability to recover meaningful model rankings depends on both the sample size and the interaction between the sampler and the exposure bias introduced during logging. Carefully designed samplers that account for item exposure and produce a small to moderate sample size, such as Skew, Popularity, and Positivity, exhibit stronger resolution, particularly under realistic constraints of logging bias and high sparsity.

### 5.2 Q2: Fidelity

To address *Q2*, we examine the fidelity of each sampling strategy, i.e. how closely the model rankings obtained from a sampled evaluation ($L_S$) align with those derived from the full reference set ($L$). We measure fidelity using Kendall's $\tau$, a rank correlation coefficient that ranges from 0 (no agreement) to 1 (perfect agreement). For each user, we compute Kendall's $\tau$ between the rankings of recommenders under the sampled and reference evaluations, and report the average across users as an overall measure of ranking agreement.

Fig. 3 presents these results. Each subplot shows average Kendall's $\tau$ (on the $y$-axis) as a function of sample size $n$ (on the $x$-axis), across different sampling strategies (plot lines). Horizontal lines indicate fixed-size samplers (Full, Exposed, and Random@$e$), which do not vary with $n$. Subplots are arranged by logging strategy (rows) and sparsity level (columns). In the first column (0% sparsity), where the logged matrix $L$ is identical to the ground-truth preference matrix $G$, the comparison is effectively between $G_S$ and $G$.

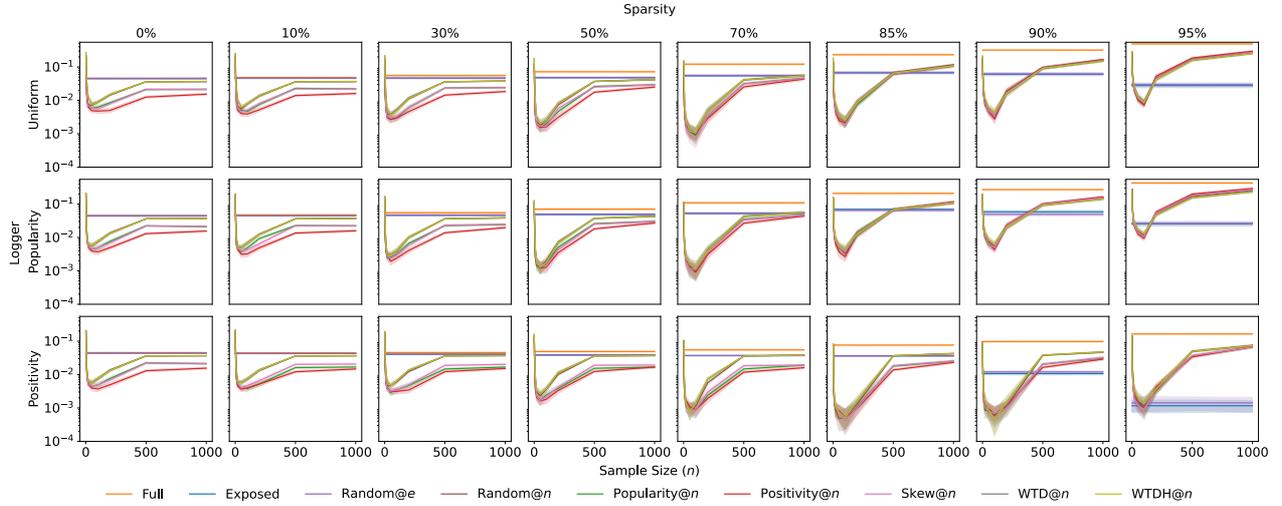

Figure 2: Average tie rate across test users (log scale, $y$-axis) as a function of sample size $n$ ($x$-axis) for different sampling strategies (plot lines), shown across logging strategies (rows) and sparsity levels (columns). A tie occurs when two recommenders achieve the same nDCG@100 for a given user. Lower tie rates indicate greater resolution. Horizontal lines represent fixed-size samplers (Full, Exposed, Random@$e$). Best viewed in color.

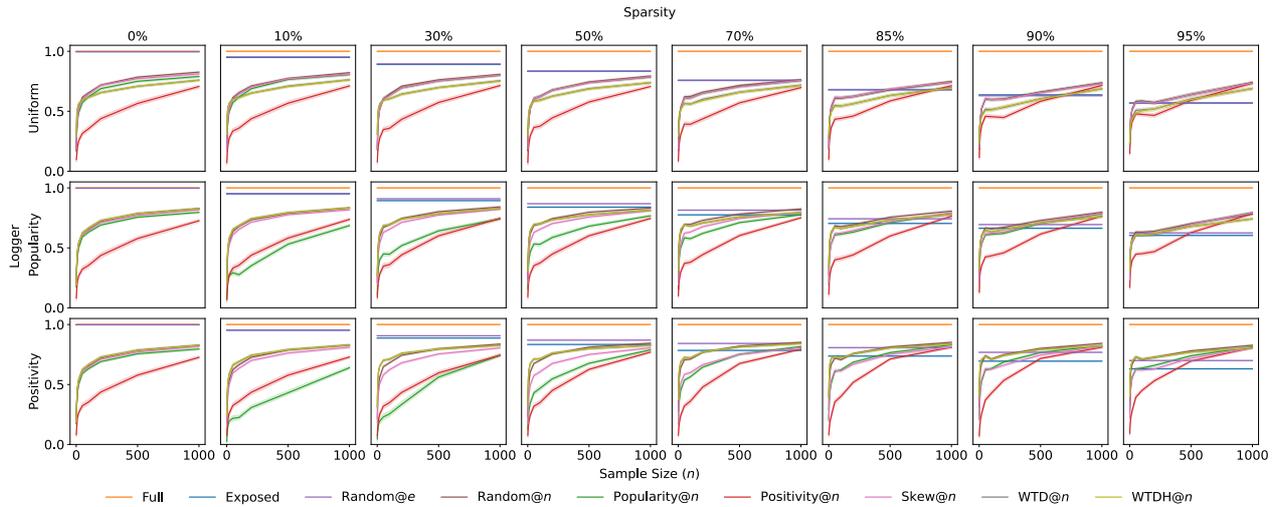

Figure 3: Average Kendall's $\tau$ correlation between model rankings under sampled evaluation ($L_S$) and their corresponding reference rankings ($L$), shown as a function of sample size $n$ ($x$-axis) for different sampling strategies (plot lines), across logging strategies (rows) and sparsity levels (columns). The $y$-axis ranges from 0 (no agreement) to 1 (perfect agreement), with higher values indicating greater fidelity. Horizontal lines indicate fixed-size samplers (Full, Exposed, Random@$e$). In the first column (0% sparsity), where $L = G$, the comparison is between $G_S$ and $G$. Best viewed in color.

As expected, fidelity improves with sample size: Kendall's $\tau$ rises quickly as $n$ increases, particularly for samplers like WTD and WTDH, which incorporate exposure or relevance signals, and Random, which assumes a uniform distribution. These methods achieve near-maximum fidelity at moderate sparsity (10%–50%) with $n \geq 200$. Skew exhibited similar performance under low sparsity conditions; however, its effectiveness declines more rapidly as sparsity increases, mainly under the positivity-biased logger. In contrast, Popularity and Positivity converge later, especially under high sparsity. Notably, fidelity depends not just on sample size, but also on which items are sampled. Although Exposed and Random@$e$

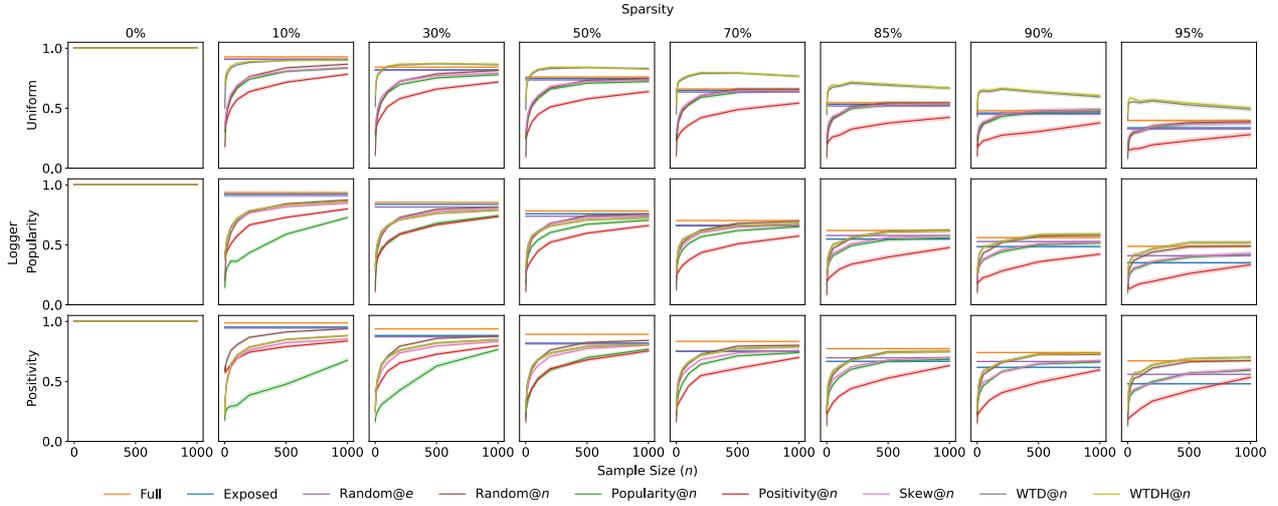

Figure 4: Average Kendall's $\tau$ correlation between model rankings under sampled evaluation from biased ($L_S$) and unbiased ($G_S$) data, with both constructed using the same sampler, shown as a function of sample size $n$ ($x$-axis) for different sampling strategies (plot lines), across logging strategies (rows) and sparsity levels (columns). The $y$-axis ranges from 0 (no agreement) to 1 (perfect agreement), with higher values indicating greater robustness to exposure bias. Horizontal lines represent fixed-size samplers (Full, Exposed, Random@$e$). Best viewed in color.

yield similar results—since both use the same number of samples per user—Exposed tends to produce lower Kendall's $\tau$, likely due to the presence of logger bias and increased sparsity. The 0% sparsity case, where $L = G$, further confirms that sampling alone can distort rankings, even in the absence of exposure bias.

Recalling *Q2*, our results suggest that fidelity is not guaranteed by sample size alone; rather, it hinges on how well a sampling strategy captures the structure and information present in the original data. Even under ideal conditions—such as 0% sparsity, where the full preference matrix is available—samplers differ markedly in their ability to preserve model rankings. This highlights that the sampling process introduces its own source of distortion, independent of exposure bias. Consequently, designing sampling protocols that align more closely with user experience can be just as important as ensuring sufficient evaluation coverage.

### 5.3 Q3: Robustness

To address *Q3*, we evaluate the robustness of each sampling strategy—that is, how sensitive model rankings are to exposure bias in the logged data. We assess this by comparing $L_S$, sampled from the biased matrix $L$, against $G_S$, sampled from the fully observed ground-truth matrix $G$, using the same sampler in both cases. Robustness is measured by Kendall's $\tau$, computed per user between rankings under $L_S$ and $G_S$, and averaged across users. Higher values indicate more consistent evaluations regardless of exposure bias.

In Fig. 4, each subplot shows average Kendall's $\tau$ (on the $y$-axis) as a function of sample size $n$ (on the $x$-axis), across different sampling strategies (plot lines). Horizontal lines represent fixed-size samplers (Full, Exposed, Random@$e$). Rows correspond to logger types and columns to sparsity levels. Since $G_S$ is fixed for a given sampler and sparsity level, all observed variation in Kendall's $\tau$ reflects how sampling interacts with bias in the logged data $L$.

From Fig. 4, we first note that, in the absence of exposure bias (first column, 0% sparsity), the comparison between $L_S$ and $G_S$ reduces to a test of internal consistency under repeated draws from the same fully observed population ($L = G$), resulting in flat lines across $n$. As in earlier sections, increasing $n$ generally improves robustness when bias is present, but the extent and consistency of this improvement vary across samplers and logging conditions. WTD, and WTDH retain high agreement between $L_S$ and $G_S$, especially with the Uniform logger, suggesting they better mitigate the distortive effects of exposure bias in scenarios with MAR data available and low sparsity. In contrast, Popularity and Positivity show limited gains, particularly under high sparsity. Full isolates the effect of exposure bias by avoiding any sampling, and performs better in the majority of cases, indicating that the choice of sampler significantly impacts evaluation results. Exposed and Random@$e$ show similar performance under low sparsity; however, as sparsity increases, Random@$e$ consistently outperforms Exposed, reinforcing once again that Exposed is highly affected by logger bias.

Recalling *Q3*, the results in this section show that most sampling strategies are indeed affected by exposure bias—but to varying degrees. While some exhibit strong robustness to bias, others diverge significantly from their unbiased counterparts, especially in sparse or skewed settings. These findings highlight that samplers cannot be evaluated in isolation from the data they operate on: robustness depends not just on the strategy, but on how that strategy interacts with the structure and bias of the logged data.

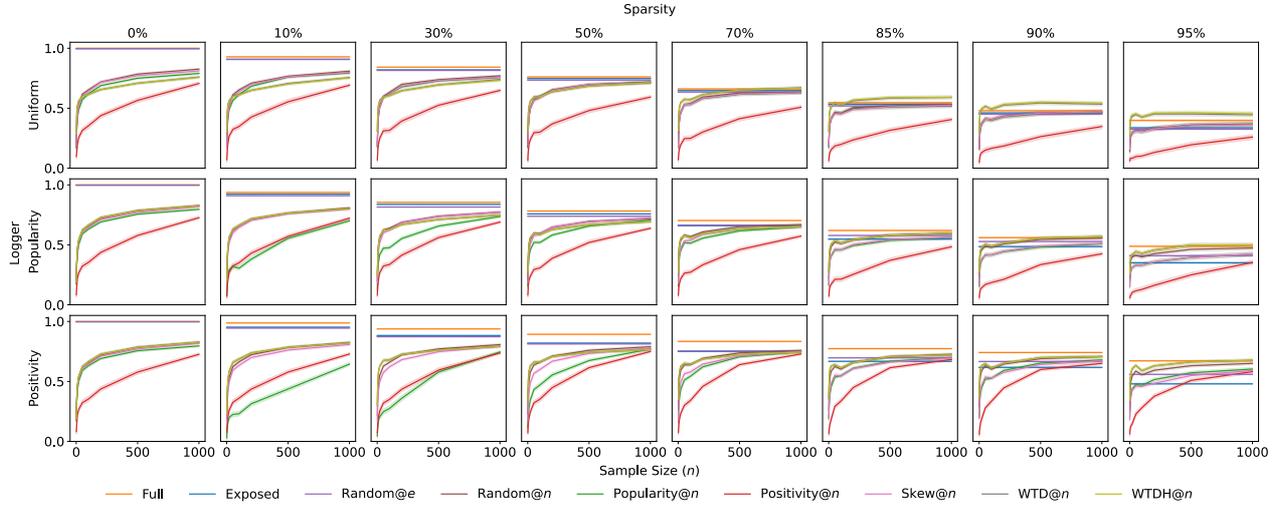

Figure 5: Average Kendall's $\tau$ correlation between model rankings under sampled evaluation from biased data ($L_S$) and the ground-truth rankings from fully observed preferences ($G$), shown as a function of sample size $n$ ($x$-axis) for different sampling strategies (plot lines), across logging strategies (rows) and sparsity levels (columns). The $y$-axis ranges from 0 (no agreement) to 1 (perfect agreement), with higher values indicating greater predictive power. Horizontal lines represent fixed-size samplers (Full, Exposed, Random@$e$). Best viewed in color.

## 5.4 Q4: Predictive Power

In the previous sections, *Q1* assessed the intrinsic resolution capacity of each sampling strategy, while *Q2* and *Q3* focused on internal agreement—either under full data or matched samplers. We now turn to *Q4*, which evaluates the predictive power of different samplers: the extent to which evaluation over biased and sampled data ($L_S$) reliably recovers the true model ranking induced by ground-truth user preferences ($G$). Predictive power is measured using Kendall's $\tau$, computed per user between model rankings from $L_S$ and $G$, and averaged across users.

As expected, predictive power improves with larger $n$, but this improvement is uneven and strongly dependent on both the sampler and the logging configuration. Skew, WTD, and WTDH again stand out, achieving strong alignment with $G$ even under moderate to high sparsity, outperforming Full in higher sparsity. These methods are more resilient to compounded distortions from logging and sampling, often reaching Kendall's $\tau$ above 0.7 at $n \geq 200$ in the 10%–50% sparsity range. Random yielded results comparable to those of weighted methods, highlighting its potential as a simple, generalizable, and easy-to-apply strategy. In contrast, Popularity and Positivity samplers exhibit much lower predictive power, especially in biased or sparse regimes. Once again, Exposed and Random@$e$ achieve similar results, reinforcing the notion that Exposed is affected by logger bias, particularly under high sparsity.

Recalling *Q4*, our results show that sampled evaluation over biased data can approximate ground-truth rankings, but only under the right conditions. Samplers like Random, WTD, WTDH, and Skew achieve high predictive power with sufficient $n$, even under moderate sparsity. In contrast, Popularity and Random@$e$ struggle, particularly in sparse or biased settings. Overall, predictive power under sampled evaluation is attainable, but requires sampling strategies that are both bias-aware and evaluation-efficient.

## 6 Conclusions

This study introduced a framework for assessing the reliability of sampling-based offline evaluation of recommender systems. Unlike prior work that targets either exposure debiasing or sampler design in isolation, our contribution lies in disentangling and analyzing their interaction. Through a meta-evaluation protocol applied to biased and unbiased data under varying sparsity and sample sizes, we assess sampling resolution, fidelity, robustness, and predictive power. Leveraging a fully observed test set, we derive ground-truth recommender model rankings to systematically quantify distortions introduced by both logging and sampling.

Our findings show that sampling strategies vary in their ability to distinguish models (*Q1*), preserve full evaluation rankings (*Q2*), remain stable under exposure bias (*Q3*), and recover ground-truth rankings (*Q4*). Strong performance depends not only on sample size, but also on sample composition and alignment with the exposure process. Bias-aware strategies such as WTD, WTDH, and Skew consistently outperform naive baselines, while Exposed emerges as a surprisingly strong contender under realistic constraints.

No single sampler excels across all criteria, requiring trade-offs: high-resolution methods may falter under bias, while robust strategies might miss fine-grained differences. Moreover, larger samples do not ensure better evaluations if item selection is biased or uninformative. These results highlight the importance of carefully designing sampling strategies, especially in sparse or biased settings, by leveraging known propensities and accounting for the interaction between logging, sparsity, and sampling.

A broader challenge going forward is the limited availability of fully observed datasets like the one used in this study, which are essential for assessing the generalizability of our findings across different domains. While such data is uncommon, it enables a level of controlled evaluation crucial for understanding the effects of exposure and sampling bias. Applying our framework to domains where dense user-item feedback can be collected—or approximated—would help validate the generality of our findings. Future work may also extend this analysis to a wider range of models, metrics, and evaluation objectives, including fairness and user trust. An especially promising direction is the development of learned sampling strategies that adaptively select evaluation candidates based on exposure patterns, model uncertainty, or past evaluation outcomes.

## Acknowledgments

This research was partially funded by the authors' individual grants from CNPq and FAPEMIG.